\documentclass[10pt,british,tightenlines,eqsecnum,floats,aps,amsmath,amssymb,nofootinbib,superscriptaddress,prd,showpacs,showkeys]{revtex4}
\usepackage{babel}
\usepackage{inputenc}
\usepackage[dvips]{graphicx}
\usepackage{amsfonts,amsmath,amssymb}

\usepackage[unicode=true,pdfusetitle,
 bookmarks=true,bookmarksnumbered=false,bookmarksopen=false,
breaklinks=false,pdfborder={0 0 1},backref=false,colorlinks=false]
 {hyperref}

\makeatletter
\@ifundefined{textcolor}{}
{%
 \definecolor{BLACK}{gray}{0}
 \definecolor{WHITE}{gray}{1}
 \definecolor{RED}{rgb}{1,0,0}
 \definecolor{GREEN}{rgb}{0,1,0}
 \definecolor{BLUE}{rgb}{0,0,1}
 \definecolor{CYAN}{cmyk}{1,0,0,0}
 \definecolor{MAGENTA}{cmyk}{0,1,0,0}
 \definecolor{YELLOW}{cmyk}{0,0,1,0}
}

\usepackage{epsfig}

\makeatother

\begin{document}

\title{\bf Two-zero textures based on $A_4$ symmetry and unimodular mixing matrix}

\author{N. Razzaghi}
\email{n.razzaghi@qiau.ac.ir}

\affiliation{Department of Physics, Qazvin Branch, Islamic Azad
University, Qazvin, Iran}

\author{S. M. M. Rasouli}

\email{mrasouli@ubi.pt}

\affiliation{Departamento de F\'{i}sica,
Centro de Matem\'{a}tica e Aplica\c{c}\~{o}es (CMA-UBI),
Universidade da Beira Interior,
 Rua Marqu\^{e}s d'Avila
e Bolama, 6200-001 Covilh\~{a}, Portugal.}


\author{P. Parada}

\email{pparada@ubi.pt}

\affiliation{Departamento de F\'{i}sica,
Centro de Matem\'{a}tica e Aplica\c{c}\~{o}es (CMA-UBI),
Universidade da Beira Interior,
 Rua Marqu\^{e}s d'Avila
e Bolama, 6200-001 Covilh\~{a}, Portugal.}

\author{P. V. Moniz}

\email{pmoniz@ubi.pt}

\affiliation{Departamento de F\'{i}sica,
Centro de Matem\'{a}tica e Aplica\c{c}\~{o}es (CMA-UBI),
Universidade da Beira Interior,
 Rua Marqu\^{e}s d'Avila
e Bolama, 6200-001 Covilh\~{a}, Portugal.}

\begin{abstract}
Applying the $A_4$ symmetry in the scenario of unimodular second
scheme of trimaximal $TM_2$ mixing matrix, where the charged
lepton mass matrix is diagonal and the nature of neutrinos are
Majorana, we investigate and analyze feasible two zeros neutrino
mass matrices.  Among the seven possible two-zero textures with
$A_4$ symmetry, we have found that only two textures, namely the
texture with $(e, e)$ and $(e,\mu)$ vanishing element of mass
matrix and its permutation, are consistent with the experimental
data in the non-perturbation method. We also obtain new
significant relations between phases of our model, namely
$\rho+\sigma=\phi\pm\pi$ and
$\sin^2{\theta_{13}}=\frac{2}{3}R_\nu$ where $R_\nu=\frac{\delta
m^2}{\Delta m^2}$. Subsequently, by admitting the experimental
ranges of $R_\nu$, we retrieve the allowed range of the unknown
phase $\phi$. Such a procedure assist us to determine the ranges
of all the neutrino observable parameters, the masses of
neutrinos, the CP-violating phases and $J$ parameter as well as to
predict the normal hierarchy for the neutrino mass. Finally, we
show that our predictions with respect to our herewith reported
specific textures are consistent
 with the corresponding data reported from
 neutrino oscillation, cosmic microwave background and
neutrinoless double beta decay experiments.

\end{abstract}


\date{25 April 2022}


\maketitle

\section{Introduction}
\label{int}

 One of the successful phenomenological neutrino mass models with
flavor symmetry, which is an appropriate framework towards
understanding the family structure of charged-lepton and of
neutrino mass matrices, is based upon the group $A_4$ \cite{mad,
madd, maddd, Alt, Hir, Fer, Mel, par, king, sta, for, lav}.
 The
$A_4$ is a symmetry group of the tetrahedron, whose introduction
was primarily motivated so that a
 tribimaximal (TBM) \cite{tbmm} mixing matrix
\cite{Fer} could be considered to explore the implications of the
mentioned charged-lepton and neutrino mass matrices. The TBM
mixing matrix is
\begin{equation}\label{etbm}
U_{TBM} =\left(\begin{array}{ccc}-\sqrt{\frac{2}{3}} & \frac{1}{\sqrt{3}} & 0\\
\frac{1}{\sqrt{6}} & \frac{1}{\sqrt{3}} &
-\frac{1}{\sqrt{2}}\\\frac{1}{\sqrt{6}} & \frac{1}{\sqrt{3}} &
\frac{1}{\sqrt{2}}\end{array}\right),
\end{equation}
where, regardless of the model, the mixing angles are
$\theta_{12}\approx {35.26}^\circ$, $\theta_{13}\approx 0$, and
$\theta_{23}\approx 45^\circ$ \cite{per}. In the last decade,
significant consequences were extracted from neutrino experiments,
such as T2K \cite{tk, tkr}, RENO \cite{ahn}, DOUBLE-CHOOZ
\cite{abe}, and DAYA-BAY \cite{fpa, daya}, which have indicated
that there are
 a nonzero mixing angle $\theta_{13}$ (at a significance level
higher than $8\sigma$) and a possible nonzero Dirac CP-violation
phase $\delta_{CP}$. Therefore, the TBM mixing matrix as above had
to be rejected \cite{cap, oli}. This consequence is in our opinion
of particular interest,
 being at the core
motivation and purpose of our paper, which we elaborate as
follows.

According to the standard parametrization, the unitary lepton
mixing matrix, which connects the neutrino mass eigenstates to
flavor eigenstates, is given by \cite{mixinga, mixingb, mixingc}

\begin{equation}\label{emixing}
U_{PMNS}=\left(\begin{array}{ccc}c_{12}c_{13} & s_{12}c_{13} & s_{13}e^{-i\delta}\\
-s_{12}c_{23}-c_{12}s_{23}s_{13}e^{i\delta} &
c_{12}c_{23}-s_{12}s_{23}s_{13}e^{i\delta} &
s_{23}c_{13}\\s_{12}s_{23}-c_{12}c_{23}s_{13}e^{i\delta}
& -c_{12}s_{23}-s_{12}c_{23}s_{13}e^{i\delta} & c_{23}c_{13}\end{array}\right)\left(\begin{array}{ccc} 1 & 0 & 0 \\
0 & e^{i\rho}& 0\\0 & 0 & e^{i\sigma}\end{array}\right),
\end{equation}
where $c_{ij}\equiv\cos\theta_{ij}\text{ and
}s_{ij}\equiv\sin\theta_{ij}$ (for $i,j=(1,2), (1,3) \text{ and }
(2,3) $); $ \delta $ is called the Dirac phase, analogous to the
CKM phase, and $\rho$, $ \sigma$ are called the Majorana phases,
which are relevant for the Majorana neutrinos. Furthermore, as
reported from experiments, the number of the known available
neutrino oscillation parameters approaches to five. In
table~\ref{h1},
 information concerning neutrino masses and mixing
provided  is summarized~\cite{expd}.

\begin{table*}[t]
\begin{center}
\begin{tabular}{ c c c }
\hline\hline Parameter &  ~~~~~~~~~~~~~~~The experimental data &~\\
\hline
~ &  ~$3\sigma$ range   & bfp $\pm1\sigma$\\
\hline\hline
$\delta m^{2}[10^{-5}eV^{2}]$ & $6.94-8.14$ & $7.30-7.77$  \\

$|\Delta m^{2}|[10^{-3}eV^{2}]$& $2.47-2.63$ & $2.52-2.53$ \\
~&$2.37-2.53$&$2.42-2.47$ \\

$\sin^{2}\theta_{12}$  & $0.271-0.369$ & $0.302-0.334$  \\

$\sin^{2}\theta_{23}$& $0.434-0.610$ & $0.560-0.588 $  \\
~&$0.433-0.608$&$0.561-0.568$\\

$\sin^{2}\theta_{13}$& $0.02000-0.02405$ & $0.02138-0.02269$\\
~&0.02018-0.02424&$0.02155-0.02289$ \\

$\delta$& $128^\circ-359^\circ$ & $172^\circ-218^\circ$ \\
~&$200^\circ-353^\circ$&$256^\circ-310^\circ$\\

\hline\hline
\end{tabular}
\caption{The experimental data associated with the neutrinos
oscillation parameters. When multiple sets of allowed ranges are
stated, the upper row and the lower row correspond to normal
hierarchy and inverted hierarchy, respectively ($\delta m^2\equiv
m_2^2-m_1^2$ and $\Delta m^2\equiv m_3^2-m_1^2 $).}\label{h1}
\end{center}
\end{table*}

In order to meet these experimental results, several models with a
discrete flavor symmetry \cite{15}, \cite{34, 35, 36}, including
an $A_4$ flavor symmetry, have been proposed \cite{15, Alt, Hir,
Fer, Mel, par, king, sta, for, lav}, \cite{37, 38, 39, 40, 41, 42,
43, 44, 45, 46, 47}. Although, the original objective of the $A_4$
models was to substantiate a TBM mixing matrix \cite{Fer}, in view
of the disagreeing observational data \cite{tk, tkr, ahn, abe,
fpa, daya}, considerable efforts,
 have been made to set
up a description conveying instead a  non-TBM mixing matrix see,
e.g., \cite{15, Alt, Hir}, \cite{Mel, par, king, sta, for, lav},
\cite{38}, \cite{39}, \cite{43, 44, 45, 46}.

Let us proceed, stating that selecting a basis where the
charged-lepton mass matrix is diagonal, a particular
representation for $A_4$ is
\cite{maddd}:
\begin{equation}\label{ema}
\cal{M}_\nu =\left(\begin{array}{ccc}a+\frac{2d}{3} & b-\frac{d}{3} & c-\frac{d}{3}\\
b-\frac{d}{3} & c+\frac{2d}{3} & a-\frac{d}{3}\\c-\frac{d}{3} &
a-\frac{d}{3}& b+\frac{2d}{3}\end{array}\right),
\end{equation}
$\cal{M}_\nu$ is invariant under the transformation $G_u$, i.e.
$G_u^T{\cal{M}_\nu}G_u=\cal{M}_\nu$, where $G_u=1-2uu^T$. The
transformation $G_u$ corresponds to the magic
symmetry\footnote{Magic symmetry is a symmetry in which the sum of
elements in either any row or any column of the neutrino mass
matrix is equal \cite{magic}.} \cite{cslam}. Thus $\cal{M}_\nu$
has also magic symmetry.
 Therefore, the mixing matrix corresponding to
$\cal{M}_\nu$ (as given by~(\ref{ema})) could be the second scheme
of trimaximal mixing\footnote{The mixing matrix corresponding to
the magic symmetry is called second scheme of trimaximal mixing.}
$(TM_2)$ \cite{TM2}, which is
\begin{equation}\label{eTM}\vspace{.2cm}
U_{TM_2} =\left(\begin{array}{ccc}\sqrt{\frac{2}{3}}\cos\theta & \frac{1}{\sqrt{3}} & \sqrt{\frac{2}{3}}\sin\theta\\
-\frac{\cos\theta
}{\sqrt{6}}+\frac{e^{-i\phi}\sin\theta}{\sqrt{2}} &
\frac{1}{\sqrt{3}} &-\frac{\sin\theta
}{\sqrt{6}}-\frac{e^{-i\phi}\cos\theta}{\sqrt{2}}\\-\frac{\cos\theta
}{\sqrt{6}}-\frac{e^{-i\phi}\sin\theta}{\sqrt{2}}
&\frac{1}{\sqrt{3}}& -\frac{\sin\theta
}{\sqrt{6}}+\frac{e^{-i\phi}\cos\theta}{\sqrt{2}}
\end{array}\right),
\end{equation}
where $\theta$ and $\phi$ are two free parameters. The first
matrix in the right hand side of~(\ref{eTM}) represents
$U_{TM_2}$, which corresponds to the magic symmetry and for the
particular case where $\theta=0$ and $\phi=0$, reduces to
$U_{TBM}$ given by~(\ref{etbm}).

 In~(\ref{ema}), by assuming the
Majorana type nature of neutrinos and an $A_4$ based symmetry for
$\cal{M}_\nu $, at least, nine free real parameters can be
obtained: three flavor mixing angles $(\theta_{13},~\theta_{12},~
\theta_{23})$, three CP violating phases $(\delta,~ \rho,~
\sigma)$ and three neutrino masses $(m_1,~m_2,~m_3)$. Additional
predictions are produced when we combine an $A_4$ symmetry with
additional constraints applied to the elements of $\cal{M}_\nu $
as  given by (\ref{ema})). The most popular constraint is the
presence of zeros in $\cal{M}_\nu $. Various phenomenological
textures, specifically texture zeros \cite{kumar,z8, z9, z10, z11,
z12, z13, z14, z15, permutation}, have been investigated in both
flavor and non-flavor basis. Such texture zeros not only cause the
number of free parameters of neutrino mass matrix to be reduced,
but also assists into establishing important relations between
mixing angles.Recently, by employing the zero texture introduced
in \cite{tex022} as well as the texture proposed in \cite{tex=22}
several parameters have been extracted as well as computed within
a novel phenomenological approach to neutrino physics.

Within the context conveyed through the preceding paragraphs, the
purpose of our paper is to investigate effects arisen from using
the two-zero textures on $\cal{M}_\nu $ given by~(\ref{ema}).
Specifically,
 assuming a Majorana\footnote{However, we should mention that
establishing the nature of neutrinos is still a controversial
subject,  which could eventually be decided by experimental
observation. In particular, by means of the nonzero magnetic
dipole moment of neutrinos ruling out Majorana neutrinos or
neutrinoless double beta decay \cite{neutrinoless} ruling out
Dirac neutrinos.} nature for neutrinos, where the charged-lepton
mass matrix is diagonal, we aim to explore the phenomenological
implications of seven two-zero textures of neutrino mass matrix
together with $A_4$ symmetry, in a scenario where $|\det U|=+1$.
This is a valuable procedure that enables to obtain a  unique
relation between the phases present in the $U_{TM_2}$ mixing
matrix, therefore allowing to extract the parameters based on a
global fit of the  neutrino oscillation data \cite{expd}. This is
the main contribution of our work. Moreover, let us also point
that it has been believed that a two-zero texture of $A_4$
symmetry can further assist into explaining a  Majorana neutrino
mass matrix. Therefore, in our paper we also proceed
systematically
 by $(i)$ employing two-zero textures of $A_4$
symmetry  and $(ii)$  comparing them with experimental data, so
that,  consequently we additionally show that only the predictions
for two-zero textures $M_\nu^{S_1}$ $(M_{ee}=M_{e\mu}=0)$, and
$M_\nu^{S_2}$ $(M_{ee}=M_{e\tau}=0)$ are consistent with the
experimental data, whilst the results of others are not.

Our paper is hence organized as follows. In section \ref{Set up},
we consider a methodology by which we reconstruct the Majorana
neutrino mass matrix with $A_4$ symmetry when the charged-lepton
mass matrix is diagonal and impose two-zero textures.
Specifically, we study all seven possible two-zero textures of
$A_4$ symmetry. In subsection \ref{S1}, we will investigate
texture $M_\nu^{S_1}$ along with an unimodular condition, by which
we obtain  constraints on Majorana phases. Moreover, we obtain
some useful relations for neutrino masses, Majorana phases and
mixing angles. Subsequently, not only we compare the consequences
of the texture $M_\nu^{S_1}$ with the recent experimental data but
also present our predictions based on the actual masses and
CP-violation parameters. In subsection \ref{S2}, we will discuss
and explore the texture $M_\nu^{S_2}$ as well as the permutation
symmetry between it and the $M_\nu^{S_1}$. Furthermore, by
applying a numerical analysis, we will discuss the predictions of
the texture $M_\nu^{S_2}$ for neutrino parameters. In subsections
\ref{S3-S6} and \ref{S7}, the other two-zero textures will be
studied. We will show that their corresponding consequences are
not in agreement with the experimental data. In section
\ref{concl}, we present our conclusions.

\section{Methodology}
\label{Set up} Assuming the Majorana nature of neutrinos, the mass
matrix $\cal{M}_\nu $ in~(\ref{ema}) is a complex symmetric
matrix. In this respect, we have shown that applying the analysis
of two-zero texture for the Majorana neutrino mass matrix based on
$A_4$ symmetry, the number of distinct cases of $\cal{M}_\nu$
in~(\ref{ema}) will be restricted to seven. In what follows,
respecting the distinguishing properties of these seven two-zero
textures, we would classify them into three categories. Here, we
first introduce them, briefly. Then, in the following subsections,
we will explain in detail how we
 can establish their corresponding models.\\

\begin{itemize}
\item {\bf Category I:}

In this category, by applying the two-zero texture of $A_4$
symmetry for $\cal{M}_\nu$ in ~(\ref{ema}), we will consider only
$M_\nu^{S_1}$ and $M_\nu^{S_2}$ textures, which are obtained by
imposing $M_{ee}=M_{e\mu}=0$ and $M_{ee}=M_{e\tau}=0$,
respectively:
\begin{equation}\label{emm110}\vspace{.2cm}
M_\nu^{S_1} = \left(\begin{array}{ccc}0 & 0 & c-\frac{d}{3}\\
0 &  c+\frac{2}{3}d &-d\\c-\frac{d}{3} &-d& d
\end{array}\right)~~~~~\text{and}~~~~~M_\nu^{S_2} = \left(\begin{array}{ccc}0 & c-\frac{d}{3} & 0\\
c-\frac{d}{3} &  d &-d\\0 &-d& c+\frac{2}{3}d
\end{array}\right).
\end{equation}
It has been shown that there is a permutation symmetry between
$M_\nu^{S_1}$ and $M_\nu^{S_2}$, such that the phenomenological
predictions of texture $M_\nu^{S_2}$ can be generated from those
of the texture $M_\nu^{S_1}$ \cite{permutation}.

\item {\bf Category II:}

In this category, we propose four two-zero textures based on $A_4$
symmetry for $\cal{M}_\nu$ in~(\ref{ema}). Namely, the textures
$M_\nu^{S_3}$, $M_\nu^{S_4}$, $M_\nu^{S_5}$ and $M_\nu^{S_6}$,
which are constructed from imposing $M_{e\mu}=M_{\mu\mu}=0$,
$M_{e\tau}=M_{\tau\tau}=0$, $M_{e\mu}=M_{\tau\tau}=0$ and
$M_{e\tau}=M_{\mu\mu}=0$, respectively:
\begin{equation}\label{emm3}\vspace{.2cm}
M_\nu^{S_3} = \left(\begin{array}{ccc}a+\frac{2}{3}d & 0 & -d\\
0 &  0 &a-\frac{d}{3}\\-d &a-\frac{d}{3}& d
\end{array}\right),~~~~~~~~~~~~~~~M_\nu^{S_4} = \left(\begin{array}{ccc}a+\frac{2}{3}d & -d & 0\\
-d &  d &a-\frac{d}{3}\\0 &a-\frac{d}{3}& 0
\end{array}\right).
\end{equation}
\begin{equation}\label{emm5}\vspace{.2cm}
M_\nu^{S_5} =\left(\begin{array}{ccc}a+\frac{2}{3}d & 0 & c-\frac{d}{3}\\
0 &  c+\frac{2}{3}d &a-\frac{d}{3}\\c-\frac{d}{3} &a-\frac{d}{3}&
0
\end{array}\right),~~~~~\text{and}~~~~~M_\nu^{S_6} =\left(\begin{array}{ccc}a+\frac{2}{3}d & c-\frac{d}{3} & 0\\
c-\frac{d}{3} & 0 &a-\frac{d}{3}\\0 &a-\frac{d}{3}& c+\frac{2}{3}d
\end{array}\right).
\end{equation}
We should note that the textures $M_\nu^{S_3}$ and $M_\nu^{S_5}$
are related through permutation symmetry to $M_\nu^{S_4}$ and
$M_\nu^{S_6}$, respectively.

\item {\bf Category III:}

Finally, another two-zero texture based on $A_4$ symmetry for
$\cal{M}_\nu$ in~(\ref{ema}), $M_\nu^{S_7}$, is obtained from
assuming $M_{\mu\mu}=M_{\tau\tau}=0$:
\begin{equation}\label{emm7}\vspace{.2cm}
M_\nu^{S_7} =\left(\begin{array}{ccc}a+\frac{2}{3}d &-d &  -d\\
-d &  0 &a-\frac{d}{3}\\ -d &a-\frac{d}{3}& 0
\end{array}\right),
\end{equation}
which has $\mu-\tau$ symmetry.
\end{itemize}

\subsection{Formalism of texture $M_\nu^{S_1}$}
\label{S1}

 In the basis where the charged lepton mass matrix is
diagonal, by employing $M_\nu^{S_1}
=U_{TM_2}^*{(M_\nu^{S_1})}_{d}U_{TM_2}^\dag$, we reorganize
the neutrino mass matrix of the texture $M_\nu^{S_1}$
as
\begin{equation}\label{emm1}\vspace{.2cm}
{M_\nu^{S_1}} = U^*_{TM_2}\left(\begin{array}{ccc}\lambda_1 & 0 & 0\\
0 &  \lambda_2 &0\\0 &0& \lambda_3
\end{array}\right)U^\dag_{TM_2},
\end{equation}
where we adopted the mixing matrix $U_{TM_2}$ given
by~(\ref{eTM}); and $\lambda_1=m_1$, $\lambda_2=e^{2i\rho}m_2$ and
$\lambda_3=e^{2i\sigma}m_3$. Now, using assumptions
$(M_\nu)_{ee}=0$ and $(M_\nu)_{e\mu}=0$, associated with the
texture $M_\nu^{S_1}$, provides two complex equations. Using the
former yields
\begin{equation}\label{emm11}\vspace{.2cm}
m_1=\left(\frac{\sin{2(\rho-\sigma)}}{2\sin{2\sigma}\cos^2{\theta}}\right)~m_2,
\end{equation}
and
\begin{equation}\label{emm12}\vspace{.2cm}
m_3=-\left(\frac{\sin{2\rho}}{2\sin{2\sigma}\sin^2{\theta}}\right)~m_2.
\end{equation}
From equations~(\ref{emm11}) and~(\ref{emm12}), we can obtain the
ratio of two neutrino mass-squared differences $R_\nu=\frac{\delta
m^2}{\Delta m^2}$ (where $\delta m^2\equiv m_2^2-m_1^2$ and
$\Delta m^2\equiv m_3^2-m_1^2 $) as
\begin{equation}\label{emm129}\vspace{.2cm}
R_\nu=\frac{-\sin^2{2(\rho-\sigma)}+4\cos^4\theta\sin^22\sigma}{\cot^4\theta\sin^22\rho-\sin^2{2(\rho-\sigma)}}.
\end{equation}
We should note that $R_\nu$ is independent of $TM_2$ phase
parameter, $\phi$.

Moreover, reemploying equations~(\ref{emm11}) and~(\ref{emm12}) gives
\begin{equation}\label{emm123}\vspace{.2cm}
\frac{m_1}{m_3}=\frac{\cot2\rho-\cot2\sigma}{\csc2\sigma\cot^2\theta}.
\end{equation}

Furthermore, complex equation $((M_\nu)_{ee}=(M_\nu)_{e\mu})=0$
yields relations
\begin{equation}\label{emm122}\vspace{.2cm}
\frac{m_1}{m_3}=\frac{\sqrt{3}\tan\theta\sin2\sigma+\sin(2\sigma+\phi)}{\sin\phi}
\end{equation}
and
\begin{equation}\label{emm125}\vspace{.2cm}
\cot2\sigma=-\left(\cos2\theta\cot\phi+\frac{\sin2\theta}{\sqrt{3}\sin\phi}\right).
\end{equation}
By inserting (\ref{emm122}) and (\ref{emm125}) into
(\ref{emm123}), we obtain
\begin{equation}\label{emm124}\vspace{.2cm}
\cot2\rho=\cot\phi+\frac{\cot\theta}{\sqrt{3}\sin\phi}.
\end{equation}
Substituting the expressions associated with two Majorana phases
from~(\ref{emm125}) and~(\ref{emm124}) into~(\ref{emm129}), we
obtain an interesting relation between $TM_2$ mixing angle
parameter $(\theta)$ and $R_\nu$:
\begin{equation}\label{emm126}\vspace{.2cm}
\sin\theta=\sqrt{R_\nu},
\end{equation}
which plays an essential role within our work, as we will now
elaborate.

Employing~(\ref{emm126}), we can rewrite relations~(\ref{emm125})
and~(\ref{emm124}) in terms of $R_\nu$ and $\phi$:
\begin{equation}\label{emm1255}\vspace{.2cm}
\tan2\sigma=-\frac{\sqrt{3}\sin\phi}{2\sqrt{R_\nu(1-R_\nu)}+\sqrt{3}(1-2R_\nu)\cos\phi}.
\end{equation}
and
\begin{equation}\label{emm1244}\vspace{.2cm}
\cot2\rho=\cot\phi+\frac{1}{\sqrt{3R_\nu}\sin\phi}.
\end{equation}
Let us also impose
$|\det U_{TM_ {2}}|=1$ \footnote{For the unitary neutrino mixing
matrix, without loss of generality, we can impose the condition
$|\det U]|=1$. This is unimodularity condition of mixing matrix
\cite{my}.}. Concretely, in our herein paper, the physics of
neutrino will be governed by the mixing matrix $U_{TM_
{2}}$~of~(\ref{eTM}) which is unitary, unimodular and rephasing
invariant. Therefore, we obtain an important relation between the
phases of $U_{TM_ {2}}$, $\phi, \rho$ and $\sigma$, which is:
\begin{equation}\label{emm10}\vspace{.2cm}
\rho+\sigma=\phi\pm n\pi,
\end{equation}
 where
$-\pi\leq\phi\leq\pi$ and $n=0,1,... $. We should note that
equation~(\ref{emm10}), which is obtained only from imposing
unimodularity condition for the mixing matrix $U_{TM_ {2}}$, is
independent of the neutrino mass zero texture.

\begin{figure}[th] \includegraphics[width=10cm]{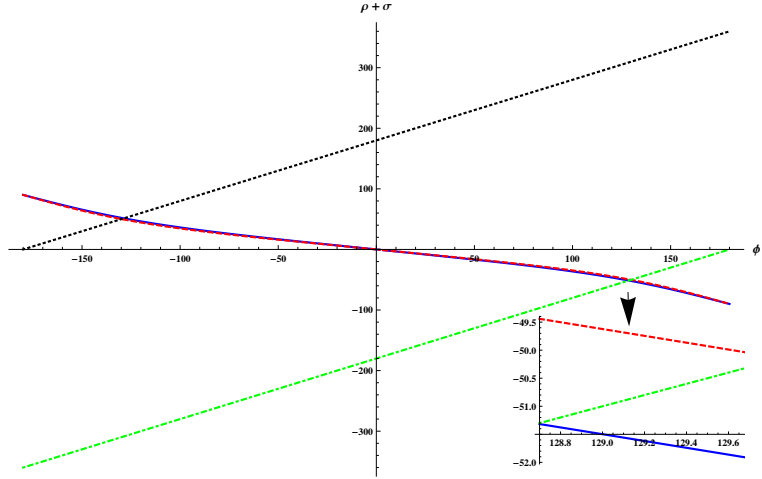}\caption{\label{fig.1}
\small{In this figure we show $\rho+\sigma$ coincide with the
lines $\phi+180^\circ$ and $\phi-180^\circ$, in which the
coincident points illustrate the allowed range of the $TM_2$ phase
parameter $(\phi)$. The black dotted line indicates the line
$\phi+180^\circ$, and the green dotdashed line indicates the line
$\phi-180^\circ$. The blue solid curve and the red dashed curve
display $\rho+\sigma$ for $R_\nu=2.64\times10^{-2}$ and
$R_\nu=3.29\times10^{-2}$, respectively. All phases and angles are
in degrees.}}
\end{figure}

Substituting Majorana phases~(\ref{emm1255}) and~(\ref{emm1244}) into~(\ref{emm10}), the most
significant consequence of our model is obtained:
\begin{equation}\label{emm1240}\vspace{.2cm}
\frac{1}{2}\tan^{-1}{\left[-\frac{\sqrt{3}\sin\phi}{2\sqrt{R_\nu(1-R_\nu)}+\sqrt{3}(1-2R_\nu)\cos\phi}\right]}
+\frac{1}{2}\cot^{-1}{\left[\cot\phi+\frac{1}{\sqrt{3R_\nu}\sin\phi}\right]}=\phi\pm
n\pi,~~~
\end{equation}
which is rewritten as a functions of only $TM_2$ phase parameter
$(\phi)$. Our endeavors have shown that equation~(\ref{emm1240})
yields acceptable results for only $n=1$, see, for instance,
figure \ref{fig.1}.

Moreover,
employing~(\ref{emm1255}),~(\ref{emm1244}),~(\ref{emm11}) and
(\ref{emm12}) as well as the definitions associated with $\delta
m^2$ and $\Delta m^2$, the neutrino masses can be expressed with
more convenient relations. More concretely, $m_1$, $m_2$ and $m_3$
are related to the unknown $TM_2$ phase parameter $(\phi)$ and the
experimental parameters $\delta m^2$ and $R_\nu$ as
\begin{eqnarray}\label{emm1266}
m_1&=&\sqrt{ \delta m^2(A-1)},\nonumber\\
m_2&=&\sqrt{\delta m^2 A},\nonumber\\
m_3&=&\sqrt{\delta m^2\left({\frac{1}{R_\nu}+A-1}\right)},
\end{eqnarray}
where
$A\equiv\left({4-4R_\nu}\right)\left({3-6R_\nu-\frac{2R_\nu\cos\phi}{\sqrt{\frac{R_\nu}{3-3R_\nu}}}}\right)^{-1}$.
Therefore, according to (\ref{emm1266}) our prediction is normal
neutrino mass hierarchy.

Furthermore, from comparing equations~(\ref{eTM}) and~(\ref{emixing}) and using~(\ref{emm126}),
we easily obtain all the mixing angles $\theta_{13}$,
$\theta_{12}$ and $\theta_{23}$ in terms of $R_\nu$ and $\phi$:
\begin{eqnarray}\label{emm1265}
\sin^2{\theta_{13}}&=&\frac{2}{3}R_\nu,\nonumber\\
\sin^2{\theta_{12}}&=&\frac{1}{3(1-\sin^2{\theta_{13}})}=\frac{1}{3-2R_\nu}.
\end{eqnarray}
According to (\ref{emm1265}), the deviation of $\theta_{12}$ from
$35^{\circ}$ depends on the value of $\theta_{13}$, where
$\theta_{13}$ depends only on $R_\nu$. Moreover, we get
\begin{equation}\label{emm1267}\vspace{.2cm}
\sin^2{\theta_{23}}=\frac{1}{2}+\frac{\sqrt{3R_\nu(1-R_\nu)}\cos\phi}{3-2R_\nu},
\end{equation}
which implies that the deviation of $\theta_{23}$ from
$45^{\circ}$ depends on the $TM_2$ phase parameter $(\phi)$. Using
\eqref{emm1267}, we can easily show
$\frac{1}{2}-\frac{\sqrt{3R_\nu(1-R_\nu)}}{3-2R_\nu}$$\leq\sin^2{\theta_{23}}\leq$$\frac{1}{2}+\frac{\sqrt{3R_\nu(1-R_\nu)}}{3-2R_\nu}$.

Moreover, $\delta\neq0$ and $\theta_{13}\neq0$ are the necessary
conditions to get CP-violation within the standard parametrization
given by (\ref{emixing}). Four independent CP-even quadratic
invariants have been known, which can conveniently be chosen as
$U^{\ast}_ {11}U_{11}$, $U^{\ast}_{13}U_{13}$, $U^{\ast}_{
21}U_{21}$ and $U^{\ast}_{23}U_{23}$. Furthermore, there is an
independent CP-odd quadratic invariants which is called Jarlskog
re-phasing invariant parameter $J$ \cite{J}. The Jarlskog
parameter is relevant to the CP violation in lepton number
conserving processes like neutrino oscillations:
\begin{equation}\label{eJ}\vspace{.2cm}
J\equiv{\cal{I}}m(U_{11}U^{\ast}_{12}U^{\ast}_{21}U_{22}).
\end{equation}
parametrization of mixing matrix $U_{PMNS}$ given by
(\ref{emixing}), the analytical expression for $J$ can be
rewritten as
\begin{equation}\label{eJ1}\vspace{.2cm}
J=\sin\delta\sin\theta_{12}\sin\theta_{23}\sin\theta_{13}\cos\theta_{12}\cos\theta_{23}\cos^2\theta_{13}.
\end{equation}
In addition, In the scheme of the $TM_2$ of mixing matrix given by
(\ref{eTM}), the analytical expression for $J$ is:
\begin{eqnarray}\label{eJ2}
J=\frac{1}{6\sqrt{3}}\cos\phi~\sin2\theta=\frac{1}{3\sqrt{3}}\cos\phi\sqrt{R_\nu(1-R_\nu)},
\end{eqnarray}
where we have used~(\ref{emm126}).

Comparing relations~(\ref{eJ1}) and~(\ref{eJ2}), as well as
reemploying (\ref{emm126}), the expression for the CP violating
Dirac phase $\delta$, in scheme of the $TM_2$ of mixing matrix,
can be written as
\begin{equation}\label{edelta1}\vspace{.2cm}
\delta=\tan^{-1}\left[\left(\frac{3-2R_\nu}{3-4R_\nu}\right)\tan\phi\right].
\end{equation}

In the present work, since we have considered massive neutrinos as
the Majorana particles, therefore, we can obtain nine physical
parameters: three neutrino masses given by~(\ref{emm1266}); three
flavor mixing angles given by~(\ref{emm1265}) and
~(\ref{emm1267}); one CP-violating Dirac phase given
by~(\ref{edelta1}); two CP-violating Majorana phases given
by~(\ref{emm1255}) and ~(\ref{emm1244}). Surprisingly, solving
equation~(\ref{emm1240}) leads to the prediction of the range of
all nine physical neutrino parameters, which were mentioned
earlier in the texture $M_\nu^{S_1}$. Let us be more precise. The
value of the $TM_2$ phase parameter $(\phi)$ can be calculated by
using two experimental data $\delta m^2$ and $\Delta m^2$, which
yield $R_\nu$. By substituting the value of
$R_\nu=(2.64-3.29)\times10^{-2}$ \cite{expd} into
equation~(\ref{emm1240}), we obtain the allowed range for the
$TM_2$ phase parameter $(\phi)$ as
\begin{equation}\label{ephas}
\phi\approx \pm(128.7^\circ-129.8^\circ).
\end{equation}
Moreover, in order to depict the allowed range of the $TM_2$ phase
parameter $(\phi)$, let us plot $\rho+\sigma$ and $\phi\pm\pi$
against $\phi$ according to~(\ref{emm1240}), see, figure
\ref{fig.1}. Obviously, the allowed range of $TM_2$ phase
parameter $(\phi)$ seen in figure \ref{fig.1} is exactly the same
as one specified in~(\ref{ephas}).

By substituting $R_\nu=(2.64-3.29)\times10^{-2}$ and $\phi$
from~(\ref{ephas}) into
relations~(\ref{emm1255}),~(\ref{emm1244}),~(\ref{emm1266}),~(\ref{emm1267}),~(\ref{eJ2})~and~(\ref{edelta1}),
not only we can obtain the ranges of the five neutrino oscillation
parameters (it is seen that these are consistent with the
experimental range of neutrino oscillation parameters in Table
[\ref{h1}]), but also we can predict the masses of the neutrinos,
the CP violation parameters, the Dirac phase $\delta$, the
Majorana phases $\rho$ and $\sigma$ and the Jarlskog invariant
parameter (which my be measured by the future neutrino
experiments).

Let us now proceed our discussions by obtaining the range of predicted values of neutrino
oscillation parameters for the texture $M_\nu^{S_1}$.

By taking $A\approx(1.2096-1.2213)$ and $\phi$ form~(\ref{ephas}),
our herein model yields
 the following values for five neutrino oscillation parameters:
\begin{eqnarray}\label{eparameter1}
\sin^2\theta_{13}&\approx&(0.01760-0.02119),\nonumber\\
\sin^2\theta_{12}&\approx&(0.3393-0.3408),\nonumber\\
\sin^2\theta_{23}&\approx&(0.4326-0.4411),\nonumber\\
\delta m^2&\approx&(6.94-8.14)\times10^{-5}eV^2,\nonumber\\
\Delta m^2&\approx&(2.47-2.63)\times10^{-3}eV^2,
\end{eqnarray}
which are in agreement with the available experimental data for
neutrino parameters in Table [\ref{h1}].

Moreover, as mentioned, our model yields the
following consequences, which may be tested by future
experiments:
\begin{eqnarray}\label{eparameter2}
m_1&\approx&(0.003918-0.004130)eV,\nonumber\\
m_2&\approx&(0.009206-0.009923)eV,\nonumber\\
m_3&\approx&(0.049912-0.051421)eV,\nonumber\\
\delta&\approx&\mp(50.84^\circ-51.80^\circ),\nonumber\\
\rho&\approx&\pm(7.46^\circ-8.40^\circ),\nonumber\\
\sigma&\approx&\mp(58.52^\circ-58.77^\circ),\nonumber\\
|J|&\approx&(0.0193-0.0220).
\end{eqnarray}
Consequently, according to the allowed ranges for the values of
three neutrino masses in~(\ref{eparameter2}), our model
successfully predicts that the neutrino mass hierarchy is normal.
Although, the corresponding relations obtained
from~(\ref{emm1266}) emphasizes enough this fact. Note that the
results of the texture $M_\nu^{S_1}$ endorse our prediction for
the neutrino mass hierarchy, which subsequently pinpoints the
corresponding relevant neutrino parameters for that mass
hierarchy.

It is worth mentioning that the texture $M_\nu^{S_1}$ together
with using $R_\nu\equiv\frac{\delta m^2}{\Delta m^2}$ assisted us
to predict all the neutrino parameters (see
relations~(\ref{eparameter1}) and~(\ref{eparameter2})), which are
in good agreement with the available experimental data. It should
be noted
 that such an ability is a distinguishing feature of the neutrino mass matrix models.

In what follows let us outline further predictions of our herein
model which can be a test on the accuracy and precision of our
predictions in ~(\ref{eparameter2})
\begin{itemize}
  \item
An important
experimental result for the sum of the three light neutrino masses
has been reported by the Planck measurements of the cosmic
microwave background \cite{planck}:
\begin{equation}\label{eplank}\vspace{.2cm}
\sum m_\nu<0.12eV \text{(Plank+WMAP+CMB+BAO)}.
\end{equation}
In our model, this quantity is predicted as $\sum
m_\nu\approx(0.063965-0.064546)$ eV, which is in agreement with
(\ref{eplank}).

\item
 Concerning the flavor eigenstates, only the
expectation values of the masses can be calculated, which is obtained from
\begin{equation}\label{enu}
\langle m_{\nu_{i} }\rangle=\sum_{j=1}^3|U_{ij}|^2|m_j|,
\end{equation}
where $i=e,~\mu,~\text{and}~\tau$. Regarding these expectation values, our predictions are:
\begin{eqnarray}\label{eparameter3}
\langle m_{\nu_e}\rangle&\approx&(0.006517-0.007065)~ eV,\nonumber\\
\langle m_{\nu_\mu}\rangle&\approx&(0.025432-0.026265)~ eV,\nonumber\\
\langle m_{\nu_\tau} \rangle&\approx&(0.031468-0.0317636)~ eV.
\end{eqnarray}

  \item
  The Majorana neutrinos can violate lepton number, for
instance, the neutrinoless double beta decay $(\beta\beta0\nu)$ was referred \cite{neutrinoless}.
 Such a process has not been
observed yet, but an upper bound has been set for the relevant
quantity, {\em i.e.}, $\langle m_{\nu_{\beta\beta} }\rangle$. For
instance, the results associated with the first phase of the
KamLAND-Zen experiment set a constraint as $\langle
m_{\nu_{\beta\beta}} \rangle < (0.061 - 0.165)~eV $ at 90 present
CL \cite{kamland}. Concerning this quantity, our model predicts:
$\langle m_{\nu_{\beta\beta}}
\rangle\approx(0.005086-0.005332)~eV$, which is consistent with
the result of kamLAND-Zen experiment.
\end{itemize}

Up to now, our herein predictions of the texture $M_\nu^{S_1}$ may suggest it
as an appropriate neutrino mass model.
Notwithstanding, it would be considered as a more successful model
if its predictions will also be supported by the cosmological
and the neutrinoless double beta decay forthcoming experiments.

\subsection{Formalism of texture $M_\nu^{S_2}$}
\label{S2}

There exists a $2-3$ permutation symmetry between textures
$M_\nu^{S_1}$ and $M_\nu^{S_2}$\footnote{There is a $2-3$
permutation symmetry which is explained that $M_\nu^{S_1}$ , and
$M_\nu^{S_2}$ are related by exchange of 2-3 rows and 2-3 columns
of neutrino mass matrix.}.Concretely, the corresponding
permutation matrix is
\begin{equation}\label{eper}\vspace{.2cm}
P_{23} = \left(\begin{array}{ccc}1 & 0 & 0\\
0 &  0 &1\\0 &1& 0
\end{array}\right).
\end{equation}
The $2-3$ permutation symmetry given by~(\ref{eper}) indicates the
following relations among their corresponding oscillation
parameters \cite{permutation}:
\begin{equation}\label{eper1}
(\theta_{13})_{s_2}=(\theta_{13})_{s_1},~~(\theta_{12})_{s_2}=(\theta_{12})_{s_1},
~~(\theta_{23})_{s_2}=90^\circ-(\theta_{23})_{s_1},~~(\delta)_{s_2}=(\delta)_{s_1}-180^\circ,
\end{equation}
Moreover, textures $M_\nu^{S_1}$ and $M_\nu^{S_2}$ have the same
eigenvalues $\lambda_i$ (for $i = 1,~2,~3$). Consequently, except
$\sin^2\theta_{23}$ and $\delta$, the other predictions for
neutrino oscillation parameters associated with the texture
$M_\nu^{S_2}$ (calculated by our model) are the same as those
predicted by the texture $M_\nu^{S_1}$ (cf subsection \ref{S1}). These
exceptions in the texture $M_\nu^{S_2}$ are:
\begin{equation}\label{es2}
(\sin^2\theta_{23})_{s_2}\approx(0.5588-0.5673)~,~~~~~~~~~~~
-(\delta)_{s_2}\approx~180^\circ\pm(50.84^\circ-51.80^\circ).
\end{equation}

\subsection{Formalism of textures $M_\nu^{S_3}$, $M_\nu^{S_4}$, $M_\nu^{S_5}$, and $M_\nu^{S_6}$}
\label{S3-S6}
\begin{figure}[th] \includegraphics[width=10cm]{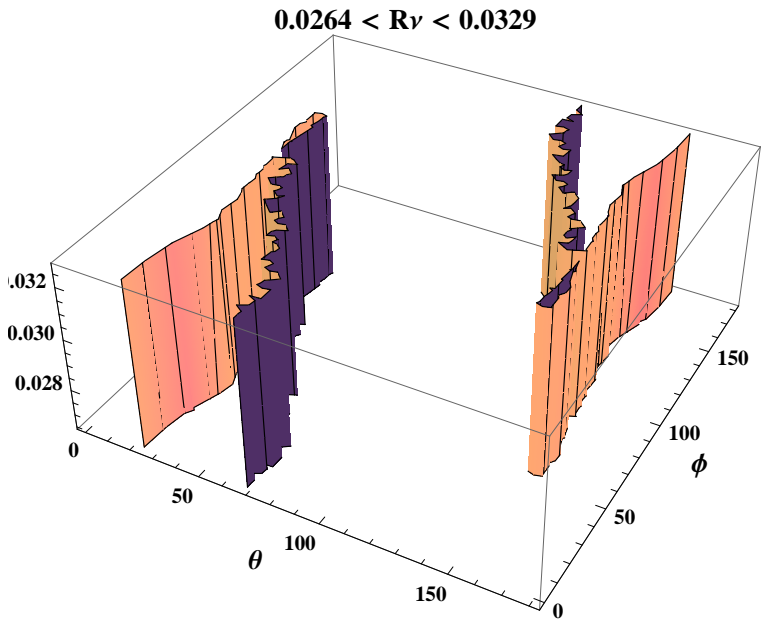}\caption{\label{fig.2}
\small{ In this figure we show the experimental value of $R_\nu$
as a function of $\theta$ and $\phi$ for the texture $M_\nu^{S_3}$
. $\theta$ and $\phi$ are in degrees.   }}
\end{figure}

The mass matrix of textures $M_\nu^{S_3}$ (see Eq.~(\ref{emm3}) has
two conditions $(M_{e\mu}=0$ and $M_{\mu\mu}=0)$, which imply the
following complex equations
\begin{equation}\label{ems31}
m_1=(1+\frac{3\sin\theta(1+
e^{2i\phi})}{\sqrt{3}e^{i\phi}\cos\theta-3\sin\theta})e^{2i\rho}m_2,
\end{equation}
and
\begin{equation}\label{ems33}
m_3=-\frac{3e^{i\phi}-\sqrt{3}\tan\theta}{3+\sqrt{3}e^{i\phi}\tan\theta
}e^{i(\phi+2\rho-2\sigma)}m_2,
\end{equation}
by which we can calculate $R_\nu$. In figure \ref{fig.2}, by depicting the
experimental value of $R_\nu$ as a function of $\theta$
and $\phi$, we have obtained the allowed range of $\theta$
around
$\theta\approx(23^\circ-70^\circ)~\text{and}~(110^\circ-157^\circ)$.
We substitute the value of $\theta$ in the expression of
$\sin^2\theta_{13}$, which, in turn, is obtained from comparing the $U_{e3}$ in
Eq.~(\ref{emixing}) with Eq.~(\ref{eTM}) as
$\sin^2\theta_{13}=\frac{2}{3}\sin^2\theta$. Finally, for the texture
$M_\nu^{S_3}$, we obtain $\sin^2\theta_{13}\approx(0.102-0.589)$,
which is inconsistent with the experimental data. From a
phenomenological point of view, the consequences associated with the
textures $M_\nu^{S_4}$ and $M_\nu^{S_3}$ are equivalent.
For the experimental values of $R_\nu$, we have shown
that these textures predict a very large values of $\theta_{13}$, which is not allowed.

Concerning the textures $M_\nu^{S_5}$ and $M_\nu^{S_6}$ (see
\ref{emm5}), we find that they predict $m_1=m_3\neq0$, which is
not allowed.

Consequently, all the textures associated with the Category II are
ruled out complectly by the experimental data listed in
table~\ref{h1}~\cite{expd}.

\subsection{Formalism of texture $M_\nu^{S_7}$}
\label{S7}

Concerning the texture $M_\nu^{S_7}$ in Eq.~(\ref{emm7}), we see
that the mass matrix has also $\mu-\tau$ symmetry. Therefore, it
implies the TBM mixing matrix with $\sin\theta_{13}=0$, which is
inconsistent with the experimental data listed in
Table~\ref{h1}~\cite{expd}.

\section{Discussion and Conclusions}
\label{concl}

In assessing neutrino physics from a phenomenological point of
view, matrix models are of particular relevance. The choice of
symmetries for the mass matrix can lead to specific states in the
mixing matrix, which may convey towards results consistent with
the corresponding experimental data. Such consequences are
significant, due to the fact that we can make additional
predictions regarding neutrinos and their flavor symmetries.

One salient feature of studying the neutrino mass matrix phenomena
is that it could, in principle, provide new keys to understand the
flavor problem; especially, its mixing matrix which has large
(mixing)angles in contrast to the quark sector. Moreover, the
disparity between the  neutrino and the charged lepton masses is
more pronounced than the corresponding features in the quark
sector. Indeed, the mass and mixing problem in the lepton sector
is a fundamental problem. Furthermore, the following important
questions should be answered by future experiments: What are the
masses of the different neutrinos? What is the nature of
neutrinos? How close to $45^{\circ}$ is $\theta_{23}$? What are
the values of three CP-violating phases associated with the
neutrino mixing matrix (i.e., the Dirac phase $\delta$ and the
Majorana phases $\rho$ and $\sigma$)?

In our work, we applied  two-zero textures within the neutrino
mass matrix with $A_4$ symmetry, along with imposing $ |\det U |=
+1$ on neutrino mixing matrix, where the charged lepton mass
matrix is diagonal and the nature of neutrinos are Majorana.
Concretely, we  have retrieved seven viable two-zero textures such
that
 the mixing matrix could be the
second scheme of trimaximal $TM_2$ mixing matrix. Assuming then
the unimodular property of the $TM_2$, we determined algebraic
 relations for Majorana phases $\rho
~\text{and}~\sigma$, together with the $TM_2$ phase parameter
$(\phi)$; cf. relation~(\ref{emm10}).

Accordingly to the physical common properties of those seven
textures, we
 classified them into  three categories. We
investigated the phenomenological properties of all these textures
and then compared them with the available  experimental data.
Among those textures, we have shown  (in the non-perturbation
method) that solely  $M_\nu^{S_1}$ and $M_\nu^{S_2}$ possess
properties that could be in agreement with the experimental data.
It is worth mentioning that applying a perturbation analysis for
the $M_\nu^{S_7}$, it may bring it to be consistent with
experimental data.
 Such an investigation has not been, however, in the scope of our present work.

 Let us be more precise. Regarding the texture $M_\nu^{S_1}$, we have shown that
 (i) $\sin\theta=\sqrt{R_\nu}$
and (ii) $\rho+\sigma=\phi\pm (+1)\pi$. This an original result
which and which leads to compute to accurate predictions for
neutrino parameters within an
 innovative as well as straightforward manner .
 Subsequently, employing the allowed ranges $R_\nu$ and $\delta\,m^2$,
 we have obtained the allowed ranges of $\phi$.
Then, we presented the predictions of our model for the values of
neutrino parameters such as mixing angles, the neutrino masses,
the expectation value of neutrino masses in the flavor bases
$i.e., (\langle m_{\nu_e}\rangle,~\langle
m_{\nu_\mu}\rangle,~\langle m_{\nu_\tau}\rangle)$, the CP
violation parameters $\delta$, $\rho$, $\sigma$, and $J$. We
emphasize that the values of all such parameters are retrieved by
merely using the allowed ranges of $R_\nu$ and $\delta\,m^2$ and
nothing else. Finally, we compared our predictions with the data
recently reported. We can conclude that there is a good agreement.
Furthermore, the predictions for  the texture $M_\nu^{S_1}$ are
also consistent with the data from the  cosmic microwave
background as well as the  neutrinoless double beta decay
experiments,  cf. ~(\ref{eparameter2}). Moreover, concerning the
texture $M_\nu^{S_1}$, we found that our prediction for neutrino
mass hierarchy is quite satisfactory.

We hope that the results of our model for the
neutrino masses, their hierarchy, CP-violation parameters
$\delta$, $\rho$, $\sigma$ and $J$ to be in agreement with the
future experiments. We have shown that there is a $2-3$
permutation symmetry between the textures. Disregarding the values
of $\theta_{23}$ and $\delta$, the mentioned symmetry yields a
similarity for the rest of predictions associated with the
textures $M_\nu^{S_1}$ and $M_\nu^{S_2}$.

In summary, applying the $A_4$ symmetry, two-zero texture
assumption, and specially including the unimodular feature of
$TM_2$ mixing matrix, we have provided the textures $M_\nu^{S_1}$
and $M_\nu^{S_2}$. We also discussed   how promising this line of
exploration can be regarding neutrino physics.

In our forthcoming investigation on neutrino physics we will be
focusing on perturbation theory to appraise states that have been
ruled out by experimental data in other frameworks. More
concretely, we will we study the $M_\nu^{S_7}$ texture in the
perturbation method, in order to assess if the corresponding will
agree with the experimental data, as we foresee it will

\section{Acknowledgments}
acknowledge the FCT grants UID-B-MAT/00212/2020 and
UID-P-MAT/00212/2020 at CMA-UBI.

\end{document}